# Orbital Occupancy and Hybridization in Strained SrVO$_3$ Epitaxial Films.

Mathieu Mirjolet,[1] Hari Babu Vasili,[2] Adrian Valadkhani,[3] José Santiso,[4] Vladislav Borisov,[5] Pierluigi Gargiani,[2] Manuel Valvidares,[2] Roser Valentí,[3] Josep Fontcuberta[1]

[1] Institut de Ciència de Materials de Barcelona (ICMAB-CSIC), Campus UAB, Bellaterra 08193, Catalonia, Spain.
[2] ALBA Synchrotron Light Source. Cerdanyola del Vallès 08290, Catalonia, Spain.
[3] Institut für Theoretische Physik, Goethe-Universität Frankfurt am Main; 60438 Frankfurt am Main, Germany
[4] Catalan Institute of Nanoscience and Nanotechnology (ICN2), CSIC, BIST, Campus UAB, Bellaterra 08193, Catalonia, Spain
[5] Department of Physics and Astronomy, Uppsala University, Box 516, Uppsala, SE-75120, Sweden

**ABSTRACT**

Oxygen packaging in transition metal oxides determines the metal-oxygen hybridization and electronic occupation at metal orbitals. Strontium vanadate (SrVO$_3$), having a single electron in a $3d$ orbital, is thought to be the simplest example of strongly correlated metallic oxides. Here, we determine the effects of epitaxial strain on the electronic properties of SrVO$_3$ thin films, where the metal-oxide sublattice is corner-connected. Using x-ray absorption and x-ray linear dichroism at the V $L_{2,3}$ and O $K$-edges, it is observed that tensile or compressive epitaxial strain change the hierarchy of orbitals within the $t_{2g}$ and $e_g$ manifolds. Data show a remarkable $2p$-$3d$ hybridization, as well as a strain-induced reordering of the V $3d(t_{2g}, e_g)$ orbitals. The latter is itself accompanied by a consequent change of hybridization that modulates the hybrid $\pi^*$ and $\sigma^*$ orbitals and the carrier population at the metal ions, challenging a rigid band picture.

**I. INTRODUCTION**

The interplay between orbital, charge and spin degrees of freedom in transition metal oxides (TMO) is at the heart of the myriad of different properties they display and it is responsible for their extreme responsivity to external stimuli. Electron density ($n$) and conduction bandwidth ($W$) are the knobs that allow fine tuning of the relative strength of these degrees of freedom. Transition metal oxides, containing $3d^n$ cations, have been much explored due to the possibility of tuning and monitoring the strength of electron-electron ($e$-$e$) correlations by increasing $n$ or reducing $W$. For instance, in the $3d^2$ compound V$_2$O$_3$, $e$-$e$ correlations open a gap in the $3d$-derived conduction band upon cooling or under pressure, and the material displays a metal-insulator transition (MIT) from a paramagnetic metal into an antiferromagnetic insulator. Electrical properties of V$_2$O$_3$ are thus understood by the presence of a Mott-Hubbard MIT, where correlations are controlled by electronic bandwidth [1,2]. However, at the MIT, there is change of crystal symmetry and the hierarchy of electronic orbitals, their electronic occupation [3] and their bandwidth [4] change. This implies also changes in hybridization between V $3d$ and O $2p$ orbitals, implying that the electron counting at $3d^n$ is not preserved and a simple $d$-orbital Mott-Hubbard description may be insufficient [5]. Similarly, in vanadium dioxide VO$_2$, changes of the $d$-$p$ hybridization in edge-connected oxygen octahedra rule the MIT [4,6]. Not surprisingly, the orbital filling and the MIT of VO$_2$ have been found to be sensitive to epitaxial strain [7].

In AMO$_3$ perovskites where the metallic $3d^n$ ions are within a corner-shared octahedra array, it is well known that lattice deformations by atomic size mismatch, epitaxial strain, etc. rule the hierarchy of $3d$ atomic orbitals. However, the relative changes of metal-oxygen hybridization by epitaxial stress remain rather unexplored. SrVO$_3$ (SVO) owing to its simple electronic configuration (V$^{4+}$, $3d^1$), its cubic structure and a relatively broad bandwidth responsible for its high electrical conductivity (with room-temperature resistivity $\rho$ = 30-50 μΩ cm) [8,9], has been the *drosophila* for research in correlated systems [10]. Here, we aim to settle if epitaxial strain acting on SVO films could induce a symmetry breaking of the $t_{2g}$(*xy, xz, yz*) orbitals significant enough to modify the 2*p*-3*d* hybridization and the charge distribution within $t_{2g}$-2*p* orbitals.

Orbital occupancy can be explored by x-ray absorption spectroscopy (*XAS*), and particularly by the x-ray linear dichroism (*XLD*) at the V $L_{2,3}$ and O *K* absorption edges. As the *XAS* intensity is proportional to the available empty states, it would allow us to probe the V 3*d* orbital occupancy as well as the V-O hybridization. Photons with an energy larger than about 515 eV can be absorbed at V $2p^{3/2}$ and V $2p^{1/2}$ core levels and the intensity of the corresponding absorption lines ($L_3$ and $L_2$, respectively) is proportional to the available lowest energy V 3*d* final states ($t_{2g}$ and $e_g$). Similarly, *XAS* absorption at O *K*-edge occurs when light is absorbed at O 1*s* core levels and electrons are excited to the lowest energy empty O 2*p* states. Observation of O *K*-edge absorption is a fingerprint of the existence of empty states at O 2*p* and thus of the covalence of the V-O bonds. The oxygen O *K*-edge occurs at about 530 eV which is only ≈ 15 eV above the V $L_2$-edge and thus the measured absorption intensity at > 530 eV contains a tail of the V $L_2$ absorption [11].

To get access to the subtle differences in the orbital occupancy of $t_{2g}$ (*xy, yz/xz*) or $e_g$ ($x^2$-$y^2$, $z^2$) states, one can collect the x-ray absorption spectra (*XAS*) for ***E***||*ab* and for ***E***||*c* (later shortened as ***E*ab** and ***E*c**, respectively), where *ab* and *c* indicate in-plane and out-of-plane x-ray electric-field ***E*** directions, respectively). The resulting dichroism (*XLD* ≈ *I*(***E*c**) – *I*(***E*ab**)) is therefore a measure of different empty states at orbitals with different in-plane (*xy* of $t_{2g}$; $x^2$-$y^2$ of $e_g$) or out-of-plane (*yz/xz* of $t_{2g}$; $z^2$ of $e_g$) symmetries. This technique has been successfully used in recent years to determine orbital occupancy within the different subsets of $e_g$ and $t_{2g}$ orbitals, in several transition metal oxides (Ti [12], V, Mn [13,14], Fe, Co [15], Ni [16–18], Cu [19,20], etc.).

Moreover, the O *K*-edge may also display a remarkable *slave* dichroism if the covalently-mixed ($p_x, p_y, p_z$) - $t_{2g}$(*xy, xz/yz*) orbitals and ($p_x, p_y, p_z$) - $e_g$($x^2$-$y^2$, $z^2$) orbitals are differently occupied. Indeed, *XLD* at O *K*-edge has been used to unravel electronic reconfigurations in manganite superlattices [21] or the nature of MIT in VO$_2$ [22].

Here, aiming at exploring and disentangling the effects of strain and covalency in SVO, films of different thicknesses have been grown on single crystalline perovskite substrates imposing different epitaxial stresses. The structural and electrical properties of the films have been inspected and their conduction band properties explored by *XAS* and *XLD* at V $L_{2,3}$ and O *K*-edges. It turns out that epitaxial strain promotes selective occupancy of V $t_{2g}$ orbitals that, in spite of the relatively weaker strength of the π*($t_{2g}$) bonds, also modulates the electron occupancy of hybridized oxygen 2*p* orbitals, where hole occupancy is also affected. Implications of these findings on the understanding of some relevant properties of metallic oxides are discussed.

## II. METHODOLOGY

We grew epitaxial [001] textured SVO films on various single-crystalline substrates by pulsed laser deposition (PLD) at a substrate temperature of 750°C. It is known that optimal transport properties of SVO films are obtained when growth is performed at the lowest pressure or using a plume-tampering Ar atmosphere [9,23]. Accordingly, we report here on films grown at the base pressure (*BP*) of the growth chamber (≈ $10^{-7}$ mbar). The number of laser pulses was varied to obtain films with nominal thickness t of 10, 20 and 70 nm, according to the growth rate calibrations. For any given number of laser pulses and pressure conditions, films on different substrates were grown simultaneously to minimize spurious thickness variations. We used single-crystalline substrates with (001)-orientation having a (pseudo)cubic lattice parameter either smaller (LaAlO$_3$ (LAO), $a_S$ = 3.791 Å), closely similar (NdGaO$_3$ (NGO), $a_S$ = 3.863 Å), and larger (SrTiO$_3$ (STO), $a_S$ = 3.905 Å) than cubic cell parameter of bulk SVO ($a_{SVO}$ = 3.842 Å). The corresponding mismatch values ($f$ = [$a_S$ − $a_{SVO}$]/$a_S$), are ($f$(LAO) = −1.37%, $f$(NGO) = +0.52%, and $f$(STO) = +1.59%). The topography of the film surface was inspected by atomic force microscopy (AFM). The surface roughness of the 10 nm films was found to be a < 0.2 nm (1x1 μm$^2$), slightly increasing up to ≈ 0.44 nm when increasing film thickness [24]. The structural characteristics of the SVO films were investigated by x-ray diffraction (XRD), using θ-2θ patterns and reciprocal space maps to determine the out-of-plane ($c$) and in-plane ($a$) cell parameters. Electrical resistivity and Hall effect measurements were performed using a PPMS (Quantum Design) under magnetic fields up to ±9 T.

We measured the XAS spectra of the samples at the V $L_{2,3}$ and O $K$-edges at 300 K and 2 K, using horizontally ($H$) or vertically ($V$) linearly polarized light and probed the *XLD* as the difference between the two light polarizations (see sketch of the measurement configuration in Figure 4(a)). The x-ray absorption was collected with $V$ (**E**||$ab$, **E**$_{ab}$) and $H$ (**E**||$bc$) polarizations, where $ab$ and $bc$ indicate the planes defined by the ($a, b, c$) crystallographic axes of the sample. For the 3$d$ orbitals, the $H$-polarized *XAS* spectrum can be different for a grazing incidence (**E**||$c$) and the normal incidence (**E**||$b$) while the $V$-polarized spectrum (**E**||$a$) remains unchanged except the probing depth. Following the common practice, we collected most of the spectra in the so-called grazing incidence with the x-ray incidence direction **k** at an angle θ = 30° with respect to the sample surface. Henceforth, we label the electric field vectors **E**$_c$ and **E**$_{ab}$ for the $H$- and $V$-polarized lights, respectively. The XAS-generated photocurrent was measured in the total electron yield (TEY) mode. *XAS* data in TEY mode can be robustly collected from 300 K to 2 K, confirming the metallic nature of the SVO films in this temperature range. Average *XAS* spectra were obtained by averaging the *XAS* intensities collected for both linear polarizations, i.e. *XAS* = [$I$(**E**$_c$) + $I$(**E**$_{ab}$)]/2. The *XLD* signal is defined as: *XLD* = $I$(**E**$_c$) − $I$(**E**$_{ab}$). All *XAS*/*XLD* experiments were performed at BL29 BOREAS beamline of ALBA synchrotron, Catalonia, Spain [25].

The electronic properties of SrVO$_3$ are calculated using density functional theory (DFT) within the generalized-gradient approximation, as available in the all-electron full-potential localized orbitals (FPLO) basis set code [26,27], and the generalized gradient approximation as exchange-correlation functional [28]. The integration in the Brillouin zone is performed using the trapezoidal method and a (20x20x20) **k**-mesh, in the -10 eV to 0 eV (corresponding to $E_F$) interval. The contributions of different orbitals to the density of states are determined based on the Wannier analysis which includes the vanadium 3$d$ and oxygen 2$p$ states around the Fermi level. The lattice parameters are taken from our measurements of SrVO$_3$ (10 nm) films grown on different substrates.

### III. RESULTS

Illustrative XRD data for SVO (10 nm) films on different substrates are shown in **Figure 1**(a) (data for all films are shown in [24]). Laue fringes in the θ-2θ patterns are well visible, assessing the film quality, and allowing to confirm the film thickness determined from growth rate, and to extract the out-of-plane lattice parameters ($c$-axis) by simulating the XRD pattern. From the fitting of θ-2θ scans and the analysis of the reciprocal space maps we deduced the cell parameters ($a, c$) and the tetragonality ratio $c/a$ shown in Figure 1(b). Dedicated reciprocal space maps were also collected to assess the absence of octahedral tilting or rotations (see [24]). It is observed that all films, except the 70 nm on LAO, have the in-plane cell parameters ($a$-axis) coinciding with those of substrates and thus these films are coherently strained on the corresponding substrates (Figure 1(a) and Figure S2 of Ref. [24]). In contrast, the reciprocal space map of the (70 nm thick) SVO//LAO film reveals the coexistence of fully strained and partially relaxed regions (Figure S2 of [24]).

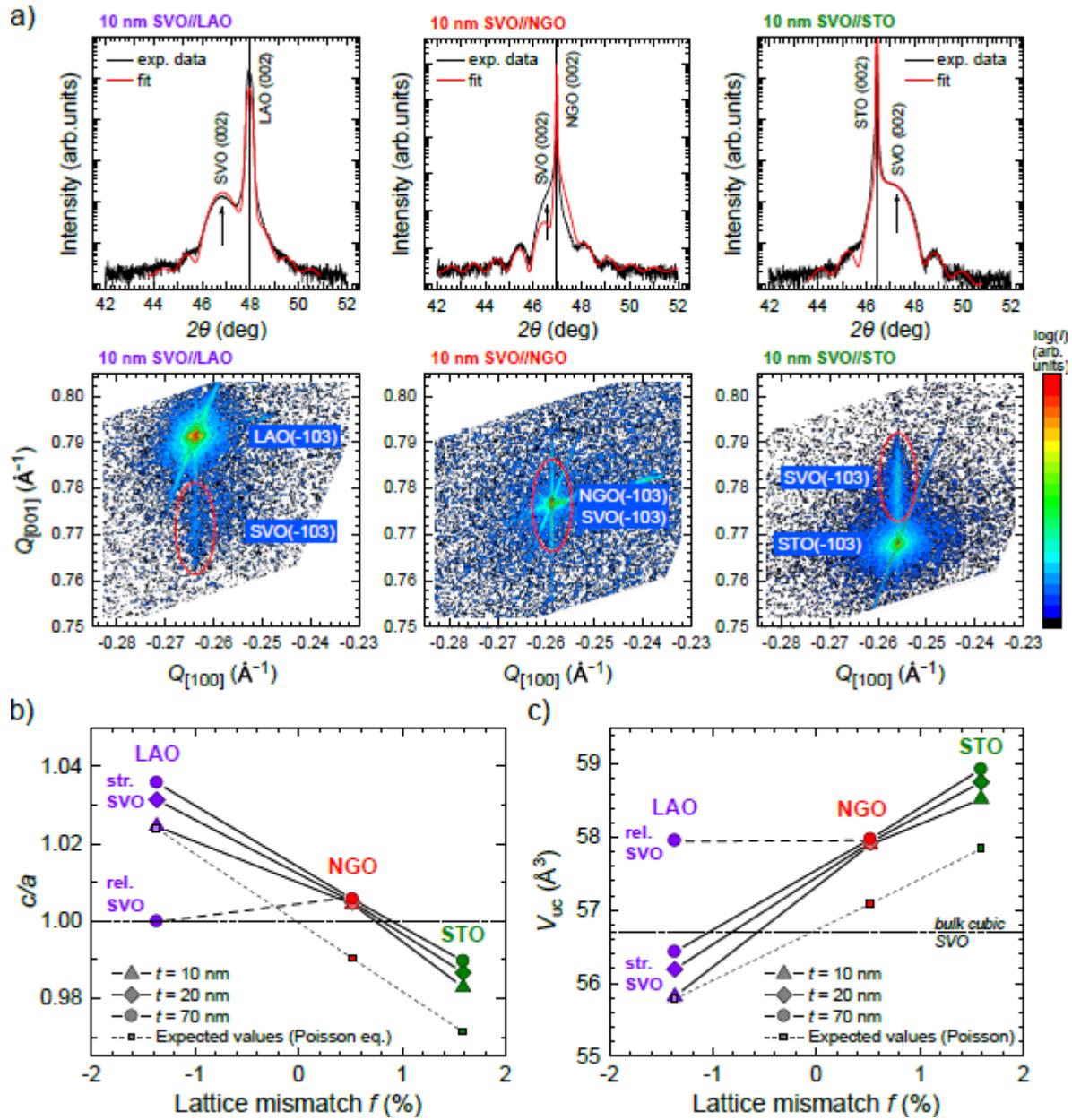

**Figure 1.** (a) Top panel: θ-2θ scans of SVO films of $t$ = 10 nm, on STO, NGO and LAO substrates. The continuous red lines are the results of the optimal simulation used to extract the $c$-axis and the film thickness. Bottom panel: the corresponding reciprocal space maps measured around the (-103) reflection. The SVO reflection is circled. (b) Experimental $c/a$ ratio for SVO films of various thicknesses (10, 20 and 70 nm), grown on LAO, NGO and STO substrates. The dashed line indicates the expected $c/a$ values for fully strained films ($a = a_S$) and $c$ is calculated using Poisson equation (with Poisson ratio $v$ = 0.28). (c) Measured unit cell volume $V_{uc}$ as a function of structural mismatch $f$. The dashed line indicates the expected unit cell volume calculated using the indicated Poisson ratio. Dotted lines in b) and c) indicate the expected $c/a$ ratio and unit cell volume of bulk SVO, respectively.

Figure 1(b) depicts the tetragonality ratio $c/a$ for all films evaluated from the extracted ($a, c$) cell parameters. It can be appreciated that the tetragonality increases from STO to NGO and to LAO. For SVO//STO films, $0.98 \leq c/a \leq 0.99$ (depending on thickness) would indicate a tensile stress compared to cubic SVO, whereas the LAO//SVO films, having $1.025 \leq c/a \leq 1.035$, would be consistent with a compressive stress. SVO//NGO films are marginally tensile stressed (1.004 ≤ $c/a$ ≤ 1.006). Overall, this is the expected structural response of a SVO film to the tensile-to-compressive film/substrate mismatch. We also include in Figure 1(b) (solid squares, dashed line) the predicted tetragonality ratio of SVO//STO and SVO//NGO and SVO//LAO films calculated using the reported Poisson ratio for SVO ($v \approx 0.28$) [29] to account for the elastic response of the SVO lattice to in-plane epitaxial strain ($\varepsilon$). For coherently grown films $\varepsilon = f$. It can be appreciated that the measured $c/a$ values of strained films display the expected dependence on epitaxial strain. However, the values of $c/a$ are found to be larger than the ones predicted using the Poisson ratio, which suggests an expansion of the $c$-axis that cannot be explained exclusively by an elastic deformation of the lattice. Observation of an anomalous expansion of out-of-plane $c$-axis in epitaxial oxide thin films, more noticeable in films on STO and NGO imposing a tensile strain, is a common finding and typically attributed to oxygen defects in the lattice, which are predicted to be more abundant in films under tensile strain [30]. Consistent with the observed partial relaxation of the compressively stressed SVO//LAO, the experimental $c/a$ values of the relaxed fraction of the film falls below the extrapolated fully-strain $c/a$ values. As shown in Figure 1(c), the measured unit cell volume ($V_{uc}$) of SVO films under tensile strain is larger than that of bulk SVO. There is a clear expansion of the unit cell with increasing tensile strain (i.e. reducing $c/a$), which indicates that point defects incorporation depends on strain, being more pronounced for tensile strain than for compressive one [30].

All SVO films reported here, including the thinnest ones ($t$ = 10 nm), are metallic (**Figure 2**(a), and Ref. [24]) with residual resistivity ratios $RRR$ (= $\rho$(300 K)/$\rho$(5 K)) ranging from 1.4-1.7 for the thinnest films (10 nm) and increasing to 1.6-2.1 for the thicker films (70 nm). In agreement with previous findings, the largest $RRR$ is obtained in films on the best matching substrate (NGO) [31]. As shown in Figure 2(b), the resistivity of the films slightly decreases upon increasing thickness. In Figure 2(c) we show the carrier density ($n$) per unit volume (1/cm$^3$) (left axis) as extracted from room-temperature Hall measurements. Carrier density values are in the (1.8-2.6)x10$^{22}$ cm$^{-3}$ range which is within the range of reported values for similar SVO films: 2.26x10$^{22}$ cm$^{-3}$ [8] and (2.0-2.3)x10$^{22}$ cm$^{-3}$ [32]. It is worth noticing that SVO films of similar thickness grown under the same nominal conditions, on LSAT and NGO, having both substrates similar mismatch, have also similar carrier concentration (2.14x10$^{22}$ cm$^{-3}$) [9,31]. It can be appreciated in Figure 2(c) that $n$ increases when increasing the tensile strain. This observation is in agreement with the observed expansion of the unit cell and the possible role of non-stoichiometric effects on this remarkable trend.

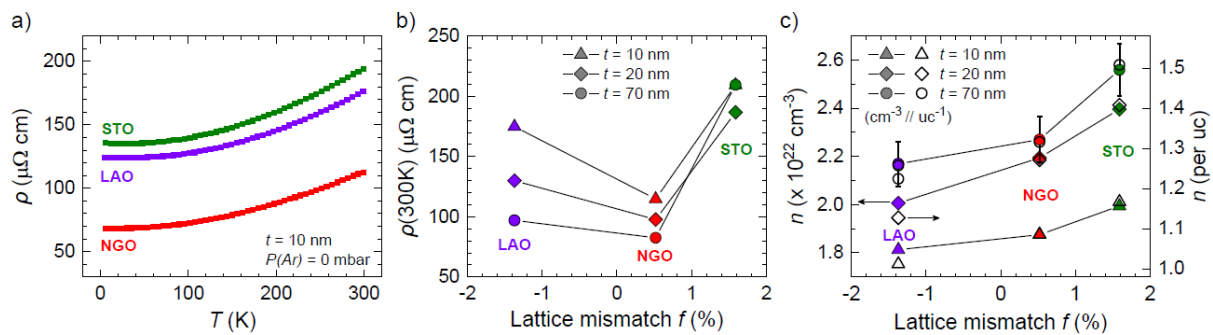

**Figure 2**. (a) Temperature dependence of the resistivity of SVO (10 nm) films grown on various substrates. (b) Dependence of the room-temperature resistivity of films of 10, 20 and 70 nm grown on substrates having different lattice mismatch ($f$) as indicated. (c) Carrier concentration ($n$) per unit volume of film (cm$^{-3}$) (left axis) and per unit cell (right axis). Error bars (shown only for the 10 nm series for the sake of clarity) are calculated assuming a maximum error of 5% in thickness determination, and neglecting any possible contribution from the substrate.

In **Figure 3**(a), we show the average *XAS* of 10 nm SVO films recorded at grazing incidence, on LAO, NGO and STO substrates, in the energy range of 510-555 eV, where the V $L_{2,3}$ absorption edge is present and followed by the O *K*-edge. The V $L_3$ (≈ 519 eV) and V $L_2$ (≈ 525.5 eV) edges are well visible but the O *K*-edge pre-peak has a slight overlap with the V $L_2$-edge and extends to a wide energy region (530-550 eV). The V $L_{2,3}$-edge is expected to differ for different valence states of $V^{m+}$ ions (e.g. $V^{3+}$, $V^{4+}$, $V^{5+}$), lowering in energy upon reducing of the valence sate, and its shape is further enriched by the presence of multiplet fine structure whose contribution largely depends on the local symmetry and the electron density [33,34]. The chemical shifts in the *XAS* $L_{2,3}$ spectra of Figure 3(a), are consistent with the V $3d^1$ electronic configuration. Moreover, the absence of a characteristic splitting occurring at the $L_2$ in V $3d^2$ systems [34], the absence of a distinctive peak at 515 eV [35] characteristic of $V^{5+}$ and the overall agreement of the shape of the *XAS* spectra with that predicted for $SrVO_3$ [36]. Moreover, the shape of the *XAS* spectra is extremely similar to that reported for isoelectronic $CaVO_3$ [37]. It follows that, within the depth probing sensitivity (≈ 5 nm) of *XAS* in TEY mode, $V^{4+}$ ($3d^1$) is the dominant formal state of the transition metal in our 10 nm SVO films. It is worth noticing that the positions of the $L_{2,3}$ peaks are preserved irrespectively on the substrate. This indicates that changes of valence state of $V^{4+}$ due to epitaxial strain, if any, are beyond experimental resolution. The XAS at $L_{2,3}$ edges of the 70 nm SVO films (Figure 3(b)) does not allow to appreciate any shift depending on the substrate. Expanded energy scale analysis, however, indicates that the $L_{2,3}$ edges of the 70 nm films are shifted by about 0.3 eV towards lower energy compared to the 10 nm films (Figure 4(b,c)). In the common rigid band picture, this would imply that $V^{m+}$ ions in the 70 nm films are somewhat reduced, say $V^{(4-\delta)+}$ ($\delta$ = 0.15-0.2), compared to the 10 nm films. This observation is in agreement with the observed slight expansion of the unit cell volume and the carrier density variation when increasing thickness observed in strained films (Figure 1(c) and Figure 2(c), respectively).

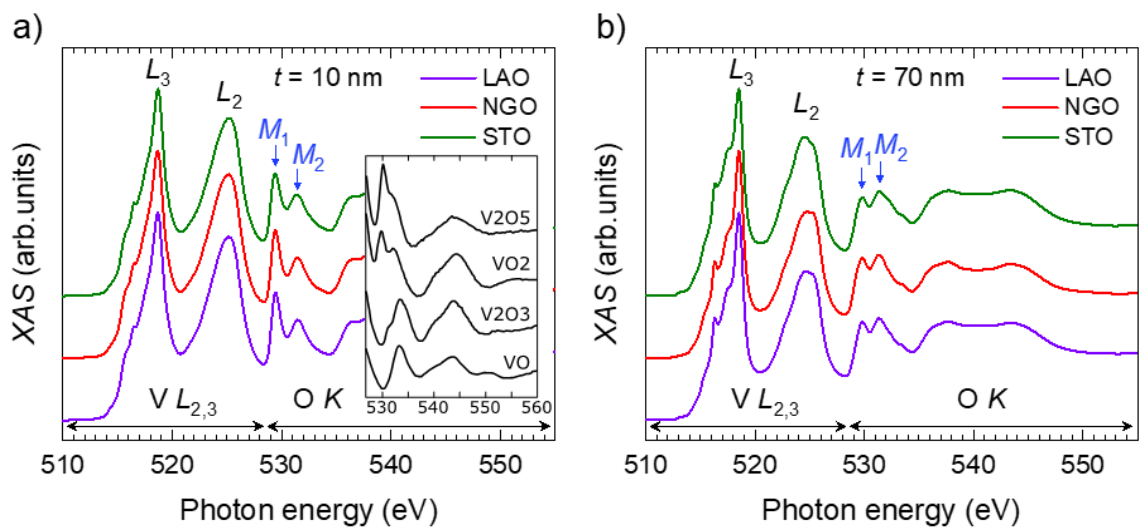

**Figure 3.** *XAS* of V $L_{2,3}$ and O *K* absorption edges recorded at room temperature, of SVO films grown on STO, NGO and LAO, and for thicknesses of: (a) 10 nm, and (b) 70 nm. The energy ranges of V $L_{2,3}$ and O *K*-edges are indicated along the *x*-axis. Arrows of $M_1$ and $M_2$ at O *K*-edge pre-peak doublet represents the main absorption lines between O 1*s* and O 2*p* states, hybridized with 3*d*-$t_{2g}$ and 3*d*-$e_g$ states. Inset in (a): illustrative O *K XAS* data for different $VO_x$ oxides (experimental data adapted from Hébert et al. [38]).

We now focus our attention to the *XAS* at O *K*-edge, associated to transitions between O 1*s* and O 2*p* orbitals, and particularly to the pre-peak doublet maxima appearing at $M_1$ (≈ 529.5 eV) and $M_2$ (≈ 531.5 eV) (indicated by arrows in Figure 3(a,b), which are known to be sensitive to oxygen contents [39]. The very presence of the $M_1$ and $M_2$ peaks indicates that empty final states are available at O 2*p* orbitals. This doublet is a fingerprint for the hybridization between O 2*p* and metal 3*d*-$t_{2g}$ and between O 2*p* and 3*d*-$e_g$ states, respectively. Correspondingly, the energy difference $\Delta M = M_2 - M_1 \approx 2$ eV, is a measure of the so-called *ligand field energy splitting* $\Delta E = E(e_g) - E(t_{2g})$ [38,40]. The relative intensity $I(M_1(t_{2g}))/I(M_2(e_g))$ is sensitive to the electronic occupancy at V 3*d* levels, which should vary according to the valence state of $V^{m+}$ ions and hybridization. We note that an accurate determination of this ratio is challenged by the presence of the tail of the vanadium $L_2$-edge; therefore, we will restrict ourselves to a qualitative analysis. For the *t* = 10 nm SVO films of Figure 3(a), the $I(M_1)/I(M_2)$ ratio is in excellent agreement with reference data for $VO_2$ oxide (as reproduced in the inset of Figure 3(a)) [38,41]. Therefore, the $I(M_1)/I(M_2)$ ratio closely matches that reported for an average V $d^1$ configuration, without perceptible changes when changing the substrates. Liberati et al., reporting *XAS* of $CaVO_3$ films grown on different substrates, obtained similar spectra to those of Figure 3(a) and concluded that $V^{4+}$ ($d^1$) oxidation state was prevalent in all films [37]. It is worth noticing that comparison of the *XAS* O *K* spectra recorded at normal and grazing incidence does not reveal any discernible chemical shift – within the experimental resolution (< 10 meV) – suggesting that, within the sensitive penetration depth, the SVO films are electronically homogeneous [24].

The O *K*-edge of the *t* = 70 nm films on various substrates (Figure 3(b)) shows a reduction of the $I(M_1)/I(M_2)$ ratio, indicating that the density of 2*p* final states has changed either due to strain-related modification of hybridization and/or a change of the amount of electrons [38]. Indeed, if the formal charge of $V^{4+}$ would reduce to $V^{3+}$, implying a higher density of electrons mostly at 2*p*-$t_{2g}$ orbitals, then the available holes at 2*p* hybrids would decrease and, correspondingly, the $M_1$ intensity would be reduced. Consistently, the Hall effect data (Figure 2(c)) signal an increase of carrier density of about 22-25%. A consistent increase of *c/a* with thickness is observed for films on LAO and STO, while thickness does not modify appreciably *c/a* in NGO (Figure 1(c)). As films on all substrates display a similar modification of the $I(M_1)/I(M_2)$ ratio, we conclude that carrier density seems to contribute to the apparent differences in O *K XAS* when changing film thickness, while it is reflected as a minor shift in V $L_{2,3}$ *XAS* as mentioned above.

We next aim at addressing if the electron distribution within the 3*d*($t_{2g}$, $e_g$) manifold is affected by substrate mismatch and film thickness. The *XAS* data in Figure 3(a) and 3(b) give a first but limited hint. Indeed, the fine structure of the V $L_3$-edge is related to final states available at $t_{2g}$ and $e_g$ orbitals. *XAS* features at the lower energy side of the $L_3$-edge were assigned to *xy* and *xz/yz* orbitals [42]. Recently, Wu et al. [36] used a configuration interaction approach to calculate *XAS* for $V^{4+}$ in octahedral coordination and noticed that these multiplet-related features are very sensitive to tetragonal deformations of the coordination $VO_6$ polyhedra. The shape of $L_3$ in our spectra (Figure 3(a,b)) closely resembles those calculated for strained SVO films [36]; however, raw V $L_3$ *XAS* data do not allow to obtain a deeper insight into electron occupancy and its dependence on substrate.

Therefore, we turn now to exploit the sensitivity of *XAS* to the polarization direction of the incoming photons to deduce *XLD* and to identify the symmetry of occupied states. As already mentioned in the experimental section, the data were collected for the light incidence direction **k** at an angle θ with the sample surface, with the electric field vector pointing along two perpendicular directions: *V* (**E**||*ab*) and *H* (**E**||*bc*), where *ab* and *bc* indicate the planes defined by the (*a*, *b*, *c*) crystallographic axes of the sample. In these polarization-dependent experiments, the spectra were collected at various θ angles (8°, 30°, 60°, and 85°) with respect to the film surface (*b*-axis) from nearly in-plane (8°) to almost normal incidence (85°). At θ = 0°, the electric field *E* of *H* polarized light is perpendicular to the sample surface, along the *c*-axis (and *E* = $E_c$); whereas, at θ = 90°, *E* is parallel to it, along the *b*-axis (*E* = $E_b$).

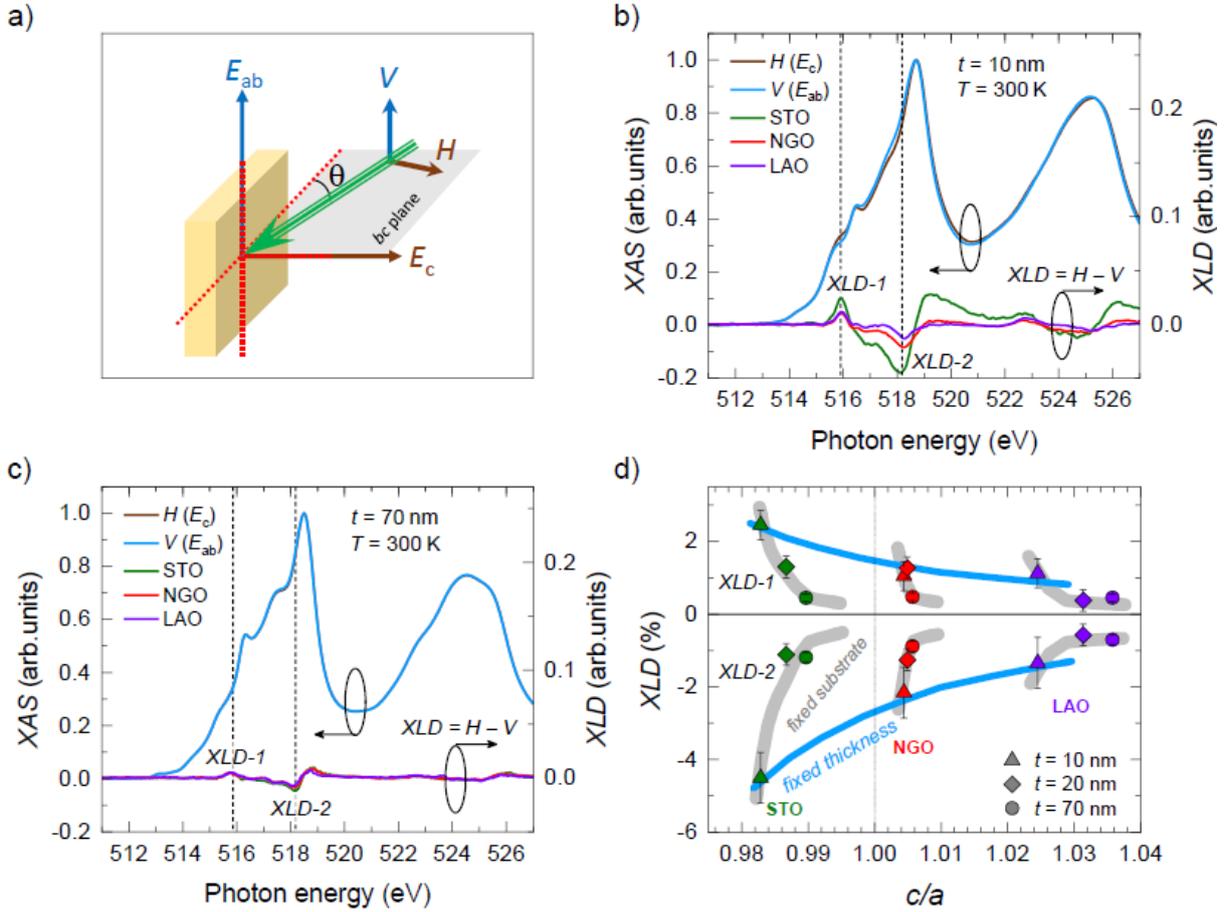

**Figure 4**. (a) Experimental arrangement for *XAS* and *XLD* measurements. (b-c) Illustrative *H*- and *V*-polarized *XAS* spectra (collected at RT and grazing indence θ = 30°) of 10 and 70 nm SVO//STO film, respectively. Bottom spectra of the panels represent the *XLD* spectra (i.e. *I*($E_c$) - *I*($E_{ab}$)) for each substrate (STO, NGO and LAO). Some reference energies, *XLD-1* and *XLD-2*, for maxima of positive and negative dichroism, respectively, are indicated by vertical dashed lines. (d) Summary of the *XLD* maxima values of SVO films for various thicknesses (10, 20, and 70 nm) grown on STO, NGO and LAO substrates.

Aetukuri et al. [7] showed that an insight into orbital occupation in $V^{4+}$ can be safely achieved by restricting the *XLD* analysis to the lowest energy excitonic part of the V $L_3$ spectra (512-516 eV, range). On the other hand, extraction of V-3*d* related *XLD* values requires appropriate normalization of the raw *XAS* spectra collected for *H* and *V* polarizations. A partial overlapping between the V $L_2$ post-edge and O *K*-edge pre-peak is known to be an issue for V *L*-edge spectroscopy, and background subtraction is prone to introduce errors in quantitative analyses [34]. Therefore, we restrict ourselves to the excitonic region and we have normalized the spectra to ≈ 528 eV, just at the $L_2$ post-edge to minimize the impact of O *K*-edge to the absorption. The spectra were further normalized to the average intensity of the V $L_3$ peak in order to assess the quantitative data analysis at V $L_3$ *XLD* peak.

In Figure 4(b) we show *XAS* at the V $L_{2,3}$-edges of 10 nm thick SVO films grown on various substrates (STO, NGO, and LAO) recorded at room temperature and θ = 30°, implying that the **E** of *V*-polarized spectra is parallel to the sample surface (**E**$_a$, commonly written as **E**$_{ab}$) and the **E** of *H*-polarized spectra is almost perpendicular to it (**E**$_c$) (Figure 4(a)). Notable differences can be observed in the raw V $L_3$ *XAS* intensities, but they are better appreciated in the corresponding dichroic *XLD* signals shown in the bottom part of Figure 4(b). Two different energies at V $L_3$ *XLD* signals are selected for the discussion, indicated by dashed vertical lines and labeled as *XLD-1* and *XLD-2*, where *XLD* displays well defined maxima for positive and negative dichroism. It is apparent that the amplitude of the corresponding *XLD-1* and *XLD-2* intensities are largest for the SVO films on STO, but reduces for the films on NGO and LAO. Data recorded at 2 K display a very similar trend [24]. *XLD* measurements have been done on the SVO films of 20 nm and 70 nm thicknesses in a similar manner. For instance, in Figure 4(c), we show the data for 70 nm SVO films (data for the 20 nm films data in Ref. [24]). It can be appreciated that *XLD* displays similar features as in the 10 nm films (Figure 4(b)) except the amplitudes at *XLD-1* and *XLD-2* are reduced with the increasing thickness. Figure 4(d) summarizes the *XLD* maxima values by displaying the amplitudes of *XLD-1* and *XLD-2* for each substrate (STO, NGO, and LAO) with different thicknesses (10, 20, and 70 nm), both parametrized by the corresponding *c/a* tetragonality ratio. Data show two main trends. First, the magnitude of *XLD-1* and *XLD-2* decreases from STO to LAO substrates, most noticeable in the thinnest films. Second, data also evidence than upon increasing film thickness and reducing the octahedral distortion |1 - *c/a*|, *XLD* progressively lowers, being the effect more remarkable in the most strained films (STO//SVO) and weaker in the partially relaxed films (SVO//LAO). Therefore, data in Figure 4(d) provides an insight on the impact of substrate and thickness on orbital occupancy in SVO films.

Figure 4(d) also contains *XLD* data of the 20 nm and 70 nm SVO films on different substrates. The same trend as in the thinnest films can be observed, with the amplitude of the *XLD* signal at *XLD-1* reducing when increasing thickness. However, we noticed above that in thicker films the *XAS* data at O *K*-edge and Hall data suggest some $V^{4+}$ reduction to $V^{3+}$, that could signal a decrease in the oxygen contents in the film, thus changing not only the electronic distribution within the 3*d* orbitals but also its density.

Therefore, we concentrate in the following on the data of the thinnest SVO (10 nm) films, where no traces of charge modification could be identified in *XAS* at V $L_{2,3}$ and O *K*-edges, as the most robust evidence of changing electron occupancy with substrate-induced stress. The *XLD* at *XLD-2* feature has its sign reversed (*XLD* < 0), with respect to *XLD-1*, and displays a mirror dependence on tetragonality ratio *c/a* and on film thickness. The presence of XLD features (*XLD-1* and *XLD-2*) of opposite sign differing by about 2.2 eV is fully consistent with calculations by Wu et al. [36].

We next focus on the sign of the dichroic signal. To minimize multiplet-configuration mixing effects [36,42,43], we restrict ourselves to the dichroic signal observed at the lowest energy range. In Figure 4(d) it is apparent that the dichroic signal at *XLD-1* is positive implying that the *XAS* intensity recorded with **E**$_{ab}$ is smaller than the **E**$_c$. In the simplest electron-hole picture, this would indicate that *xy* orbitals are more occupied than *xz/yz*. Data show that this orbital polarization is gradually reduced from STO to NGO to LAO, although *XLD* (*XLD-1*) remains positive for all films (*t* = 10 nm). Accordingly, the *xy* orbitals are most favorably occupied in all films, irrespectively of *c/a* > 1 or *c/a* < 1. *XLD* measurements were recorded at different angles (8°, 30°, 60°, and 85°) for the thinnest films (10 nm), confirming the systematic variation of the *XLD* signal with substrate [24].

As mentioned, electronic occupancy at metal $t_{2g}$ orbitals should have its fingerprint on the O *K*-edge *XAS* and the *XLD* at O *K*-edge. Accordingly, *XLD* data at O *K*-edge has been determined as for V $L_{2,3}$. In **Figure 5** we show the *XAS* spectra in the O *K* region of 10 nm SVO films grown on the different substrates, collected at the grazing incidence (θ = 30°) with the $E_c$ and $E_{ab}$ polarizations as indicated.

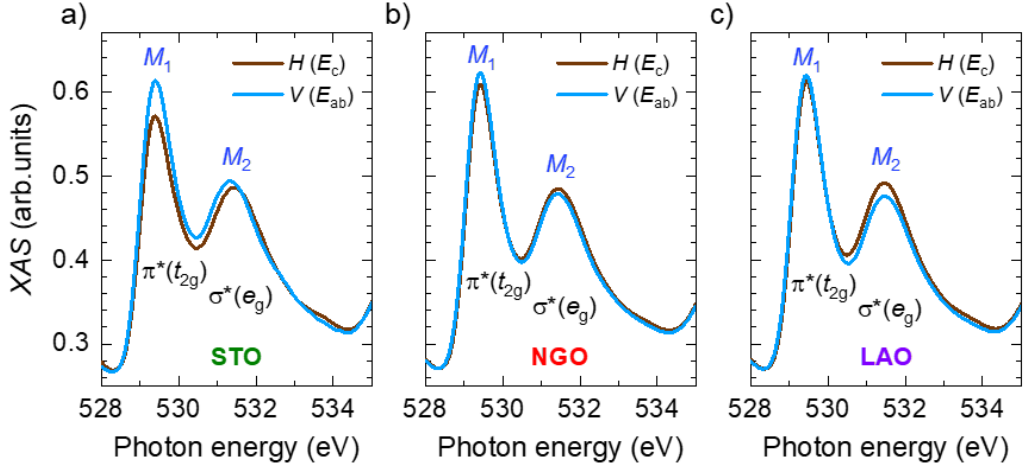

**Figure 5.** O *K*-edge *XAS* pre-peak of 10 nm SVO films grown on: (a) STO, (b) NGO, and (c) LAO. A significantly larger dichroism at $t_{2g}$ peak for STO than for NGO/LAO and the reverse at $e_g$ peak reflect changes of hybridization with strain. Data were recorded at room temperature.

In order to analyze the *XAS* data in Figure 5 we recall that the $M_1$ and $M_2$ peaks correspond to available states at O 2*p* orbitals hybridized with ligand field split V 3*d* states, resulting in $\pi^*(t_{2g})$ and $\sigma^*(e_g)$ orbitals. The intensity of the $\pi^*(t_{2g})$ absorption peak is considerably larger than the one of the $\sigma^*(e_g)$ peak: $I(\pi^*(t_{2g}))/I(\sigma^*(e_g)) \approx 1.25$ (**Figure 6**, right axis). This difference originates from the larger multiplicity of the $t_{2g}$ orbitals compared to $e_g$ ones (3/2 = 1.5) modulated by the distinct hybridization of $\pi^*(t_{2g})$ and $\sigma^*(e_g)$ orbitals. It can be also appreciated in Figure 5 that the energy difference between $\Delta E_{CF} = E(\sigma^*(e_g)) - E(\pi^*(t_{2g}))$ is of about 2.0 eV as commonly found for early transition metal oxides [40,44], and virtually insensitive to the substrate. Next we focus on the O *K* XLD signal defined as the *XAS* intensity at the corresponding $I(\pi^*(t_{2g}))$ and $I(\sigma^*(e_g))$ maxima recorded using $I(E_{ab})$ and $I(E_c)$ (*XLD* = $I(E_c) - I(E_{ab})$). In Figure 6 (left axis) we show the *XLD* values for films on various substrates.

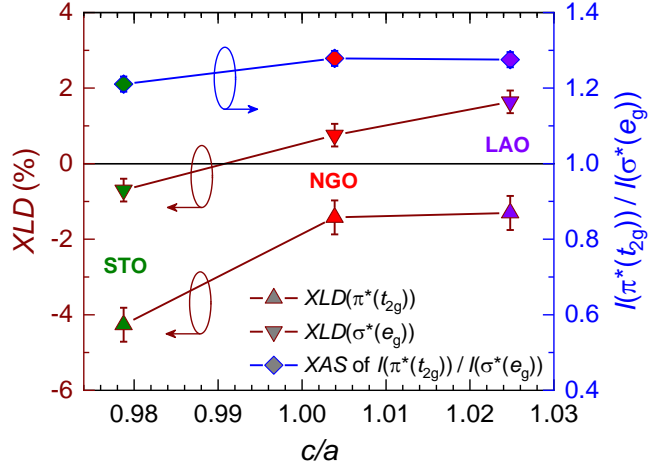

**Figure 6.** Left axis: *XLD* at O *K*-edge for SVO films (10 nm thick) on substrates imposing different tetragonality ratios (*c/a*). Square symbols indicate the *XLD* at $\pi^*(t_{2g})$ and $\sigma^*(e_g)$ absorption peaks, as indicated. Right axis: $I(\pi^*(t_{2g}))/I(\sigma^*(e_g))$ intensity ratios. Data were recorded at room temperature.

It is clear in Figure 6 (and data in Figure 5(a)) that for the tensile strained SVO//STO film (smallest *c/a* ratio), the *XAS* intensity $I(\pi^*(t_{2g}))$ is larger for $E_{ab}$ than for $E_c$, and accordingly XLD($\pi^*(t_{2g})$) < 0. The observation that $XLD(\pi^*(t_{2g})) < 0$ indicates a higher concentration of holes at $(p_x + p_y)$ orbitals of the $(p_x + p_y)$-$d_{xy}$ hybrid. In other words, strain modifies the $\pi^*(t_{2g})$ *p-d* hybridization, driving charge (for *c/a* < 1) from the $p_x + p_y$ orbitals towards the metal. An analogous reasoning accounts for the observed reduction of $XLD(\pi^*(t_{2g}))$ when increasing *c/a*. Similarly, the dependence of $XLD(\sigma^*(e_g))$ on *c/a* also reflects the corresponding changes of $\sigma^*(e_g)$ hybridization.

These observations can be rationalized on the basis of strain modification of 2*p*-3*d* hybridization and subsequent changes in the electron occupancy at 2*p*-3*d* hybridized orbitals. We notice that in an octahedral VO$_6$ environment, symmetry arguments dictate that, focusing on the VO$_2$ plane of SVO structure, $d_{xy}$ hybridizes with $(p_x + p_y)$; $d_{xz}$ hybridizes with $(p_x + p_z)$ and $d_{yz}$ hybridizes with $(p_y + p_z)$, whereas $d_{x2-y2}$ hybridizes with $(p_x + p_y)$ and $d_{z2}$ with $p_z$, as illustrated in the **Figure 7** (central panel) and Figure 7(a-b).

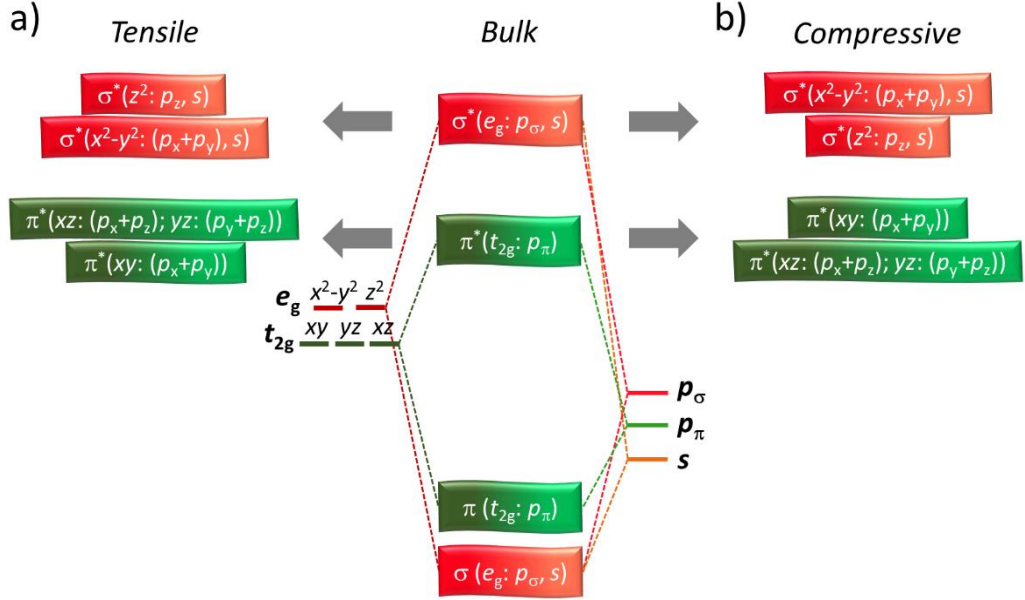

**Figure 7.** Central panel: Energy-band diagram for SrVO$_3$ thin films. The hybridized ($\sigma$, $\sigma^*$) and ($\pi$, $\pi^*$) orbitals, and their parentage, are indicated. The symmetry broken hybridized orbitals under the effect of: (a) tensile strain, and (b) compressive strain.

Under tensile strain (SVO//STO), $d_{xy}$ orbitals are pushed down as observed by *XLD*(V $L_{2,3}$). Consequently, the hybridized orbital is shifted down (Figure 7(a)) and the ($p_x$, $p_y$) orbitals which are hybridized with $d_{xy}$, are electron-depleted by charge transfer to the metal. Accordingly, *XAS* at O *K*-edge should be larger for $E_{ab}$ (more holes available) than for $E_c$, and thus *XLD*($\pi^*(t_{2g})$) should be negative (< 0), as we observed. Similarly, when SVO films are under compressive stress (SVO//LAO), $d_{xz,yz}$ orbitals are shifted down in energy as observed by *XLD*(V $L_{2,3}$). Therefore, the hybridization of these orbitals with the corresponding $p_z$ orbitals (Figure 7(b)) implies that the $p_z$ orbitals are electron-depleted (hole-rich) and the correspondingly *XLD*($\sigma^*(e_g)$) is positive (> 0), as we experimentally observed (Figure 6).

In order to analyze the role of tetragonal distortion on the electronic structure of SVO, we performed first principles DFT calculations. Of interest here is the integrated partial density of states (*IDOS*) associated to oxygen ($p_x$, $p_y$, $p_z$) and vanadium 3*d* ($t_{2g}$, $e_g$) orbitals and to disclose how their relative weight evolve with *c/a*.

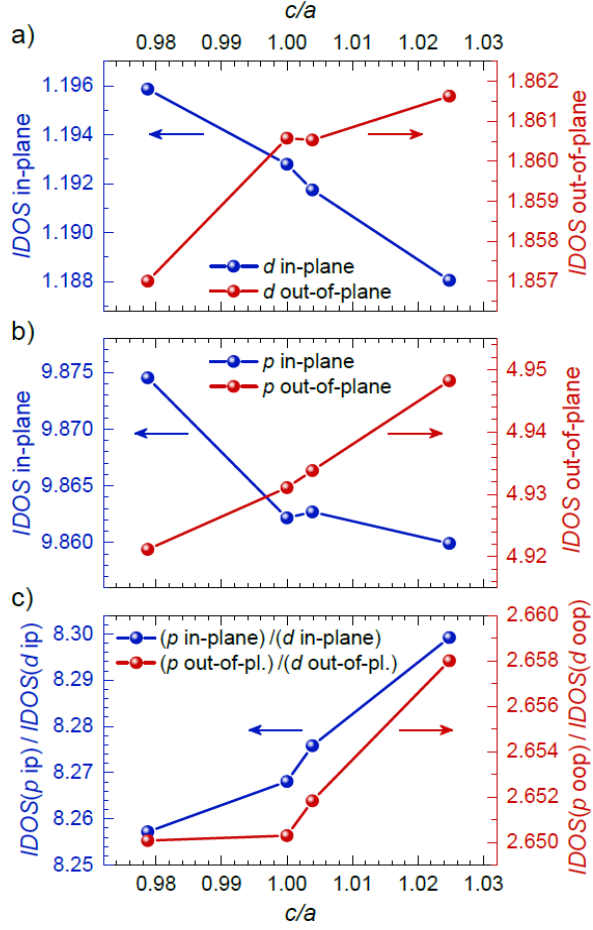

**Figure 8.** (a) Sum of the integrated partial density of states (*IDOS*) of in-plane (left axis) and out-of-plane (right axis) 3*d* orbitals. (b) Similar data for in-plane (left axis) and out-of-plane (right axis) 2*p* orbitals. (c) Ratio between the total density of states of 2*p* and 3*d* character *IDOS*(2*p*)/*IDOS*(3*d*) in-plane (left) and out-of-plane (right).

In **Figure 8**(a) we show the sum of the integrated density of states IDOS of in-plane ($x^2$-$y^2$, $xy$) and out-of-plane ($z^2$, $xz$, $yz$) orbitals of 3*d*-$e_g$ and 3*d*-$t_{2g}$ manifolds, as a function of the tetragonal distortion *c/a* determined in SVO films on STO, NGO and LAO (*c/a* = 0.978, 1.003, 1.024, respectively). We also include in Figure 8(a) the data for a cubic SVO of cell parameter *a* = *c* = 3.86 Å, corresponding to the unit cell parameter of an unstrained cubic SVO film of volume 57.5 Å$^3$ (Figure 1(c)). *IDOS* plots for every individual orbital are included in Ref. [24]. Data in Figure 8(a) clearly show that in-plane orbitals are stabilized under tensile strain (*c/a* < 1); the opposite trend is observed for out-of-plane orbitals. This is agreement with data in Figure 4(d). The same trend can be appreciated in Figure 8(b) where we show *IDOS* of in-plane and out-of-plane 2*p* orbitals. Therefore, *p-d* hybridized in-plane orbitals move in unison under tetragonal cell distortion, and similarly the out-of-plane hybridized orbitals. However, the relative *p/d* relative weight in hybrid orbitals is not preserved when changing *c/a*, as clearly indicated by data in Figure 8(c), where the *IDOS* ratios (*IDOS*(*p*)/*IDOS*(*d*)) for in-plane (left) and out-of-plane (right) orbitals are depicted (see also [24]). This implies a charge redistribution among V-O bonds under strain. Indeed, the 2*p* orbitals become progressively more occupied when increasing *c/a*. Oppositely, for *c/a* < 1, the ratio [*IDOS*(*p*, in-plane)/*IDOS*(*d*, in-plane)] lowers compared to its value for *c/a* = 1. This implies that the in-plane orbitals of 3*d* character are pushed down compared to the corresponding hybridized 2*p* orbitals, which thus have a relatively lower *IDOS*, as argued above (charge balancing sketches are shown in [24]). Correspondingly, the available states at 2$p$ in-plane orbitals become larger. This accounts for the observed $XLD(\pi^*(t_{2g})) < 0$ observed for $c/a < 1$ as shown in Figure 6. A similar reasoning accounts for the observed $XLD$ variation for $c/a > 1$.

## IV. SUMMARY AND CONCLUSIONS

In summary, bulk SrVO$_3$ is cubic, but when SVO films are grown on substrates having different structural mismatch with SVO, epitaxial growth imposes compressive or tensile strain on the film structure and its tetragonality ratio can be varied from $c/a > 1$ to $c/a < 1$ depending on the substrate used and the film thickness. *XAS* at V $L_{2,3}$ and O *K*-edges of the thinnest films (10 nm) display almost identical features fully consistent with the expected 3$d^1$ V$^{4+}$ electronic configuration of this oxide. *XLD* is well visible at $L_{2,3}$-edges, indicating that the 3$d$-$t_{2g}$ orbitals are not degenerate but signaling a clear hierarchy of (*xy*, *xz*, *yz*) orbitals that gradually varies with the epitaxial strain. In films having an in-plane tensile strain (SVO//STO), the in-plane *xy* orbitals are preferentially occupied by electrons, gradually levelling out in films under compressive strain. *XAS* at O *K*-edge provides a clear evidence of a relevant 2$p$-3$d$ hybridization. *XLD* at O *K*-edge indicates that hybrid $\pi^*$ orbitals in epitaxially tensile strained films, having an in-plane symmetry, have a hole-density that decreases in compressive strained films. The consistent variation of occupancy in 2$p$ and 3$d$ orbitals with strain shows that substrate-induced symmetry breaking modulates orbital occupancy at the metal site but also the metal-oxygen hybridization. It follows that charge density at the metal site is not preserved under strain but redistributes within the hybridized bonds. However, whereas in the case of edge-shared coordination polyhedral (VO$_2$ case) changes of hybridization with temperature or strain are strong enough to promote a metal-insulator transition, in the corner-connected octahedral networks, strain slightly modifies the electrical conductivity but SVO films (at least ≥ 10 nm) remain metallic. Still, the rigid band image of an electron redistribution restricted within the 3$d$-$t_{2g}$ manifold and dictated by strain does not hold in the simplest 3$d^1$ perovskite, with corner-sharing octahedral network, but hybridization plays a relevant role.

**Acknowledgements**

Financial support from the Spanish Ministry of Science, Innovation and Universities, through the "Severo Ochoa" Programme for Centres of Excellence in R&D (SEV-2015-0496) and the MAT2017-85232-R (AEI/FEDER, EU), and from Generalitat de Catalunya (2017 SGR 1377) is acknowledged. The work of M.M. has been done as a part of the PhD program in Physics at Universitat Autònoma de Barcelona, and was financially supported by the Spanish Ministry of Science, Innovation and Universities (BES-2015-075223). A.V. and R.V. acknowledge support by the Deutsche Forschungsgemeinschaft (DFG, German Research Foundation) for funding through TRR 288 - 422213477 (project B05). J.S. acknowledges support from project n° 2017 SGR 579 from AGAUR, Generalitat de Catalunya.

# Orbital Occupancy and Hybridization in Strained SrVO₃ Epitaxial Films.

Mathieu Mirjolet,[1] Hari Babu Vasili,[2] Adrian Valadkhani,[3] José Santiso,[4] Vladislav Borisov,[5] Pierluigi Gargiani,[2] Manuel Valvidares,[2] Roser Valentí,[3] Josep Fontcuberta[1]

[1] Institut de Ciència de Materials de Barcelona (ICMAB-CSIC), Campus UAB, Bellaterra 08193, Catalonia, Spain.
[2] ALBA Synchrotron Light Source. Cerdanyola del Vallès 08290, Catalonia, Spain.
[3] Institut für Theoretische Physik, Goethe-Universität Frankfurt am Main; 60438 Frankfurt am Main, Germany
[4] Catalan Institute of Nanoscience and Nanotechnology (ICN2), CSIC, BIST, Campus UAB, Bellaterra 08193, Catalonia, Spain
[5] Department of Physics and Astronomy, Uppsala University, Box 516, Uppsala, SE-75120, Sweden

**SUPPLEMENTARY INFORMATION**

**Supplementary Information S1**

In Figure S1, we show AFM topographic images of the thickest SVO films (70 nm) grown on different substrates. A morphology of multi-islands terraces one-unit cell (u.c) high in 3D steps (≈ 0.4 nm) can be observed, reminiscent of so-called "wedding-cake" structures, as commonly observed in thin films of oxides [45,46] or metals [47] and commonly attributed to the Ehrlich–Schwoebel barrier (ES) energy barrier for adatoms diffusion [48,49], that promotes multi-terrace island formation. As indicated in Figure S1, the roughness of these thick films is of about one-unit cell (*rms* ≈ 0.4–0.5 nm for an image size up to 1 x 1 μm²) depending on the substrate.

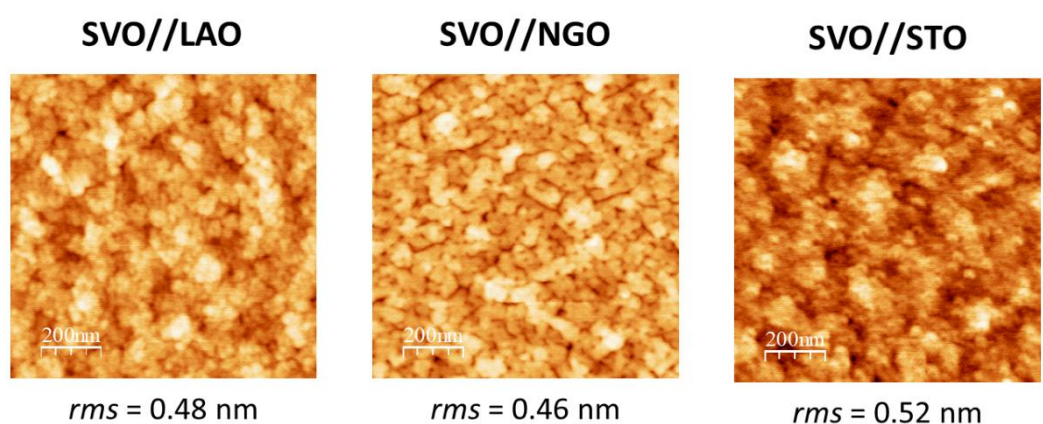

**Figure S1:** AFM topographic images of SVO films, about 70 nm thick, deposited on LAO, NGO and STO.

**Supplementary Information S2**

In Figure S2(a) we show the $\theta$–$2\theta$ scans of the SVO films of $t$ = 20/70 nm, grown on LAO, NGO and STO. The continuous line through the data are the results of the optimal simulation used to extract the $c$-axis and the film thickness. The corresponding reciprocal maps measured around the (-103) reflection are shown in Figure S2(b). For SVO//STO, the film reflection is clearly visible above the substrate one, which confirms fully tensile strained films. For SVO//NGO, film and substrate reflections are perfectly superposed due to close structural mismatch. For SVO//LAO, the film peak is located below the substrate one. In the thickest film one small portion of the peak is vertically aligned with the substrate and the rest shows strain relaxation. However, in thinner SVO//LAO samples, the film seems almost fully strained with fewer signs of relaxation.

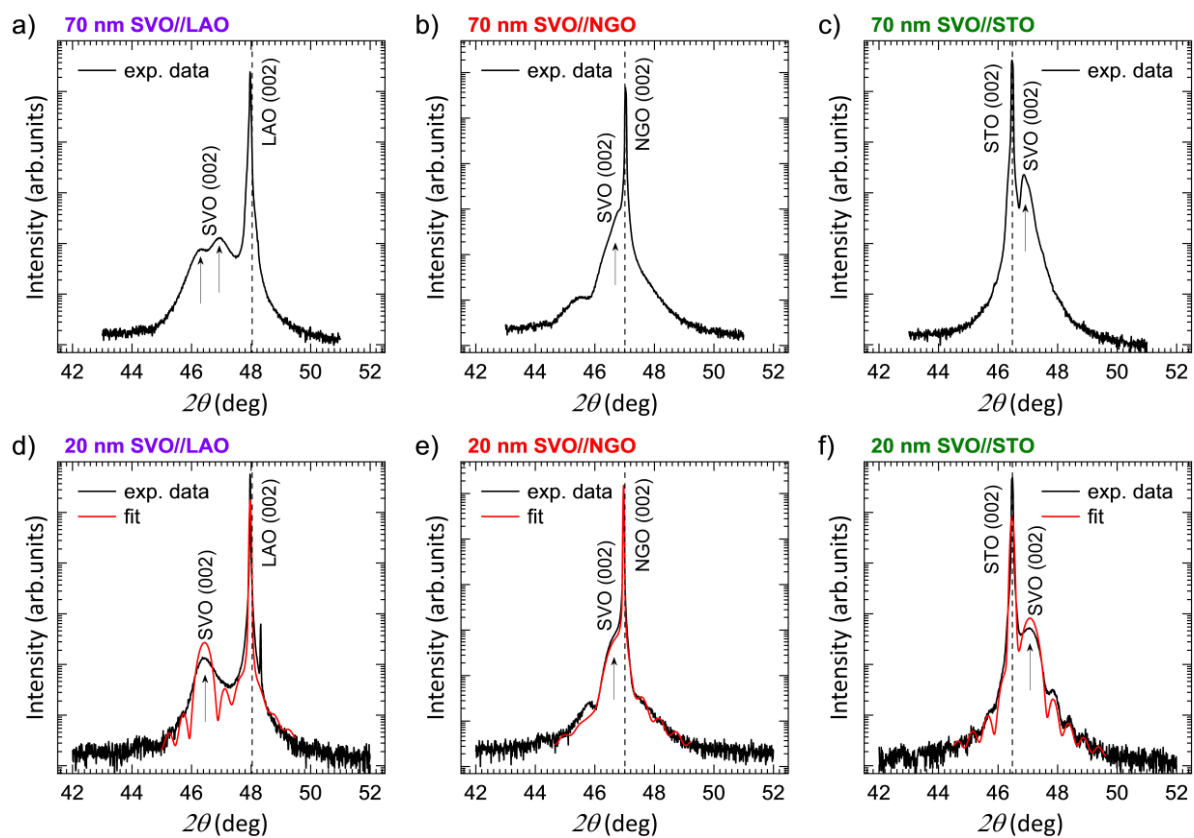

**Figure S2(a):** $\theta$–$2\theta$ scans of the SVO films of $t$ = 20/70 nm, grown on STO, NGO and LAO. The continuous red line through the data are the results of the optimal simulation used to extract the $c$-axis and the film thickness.

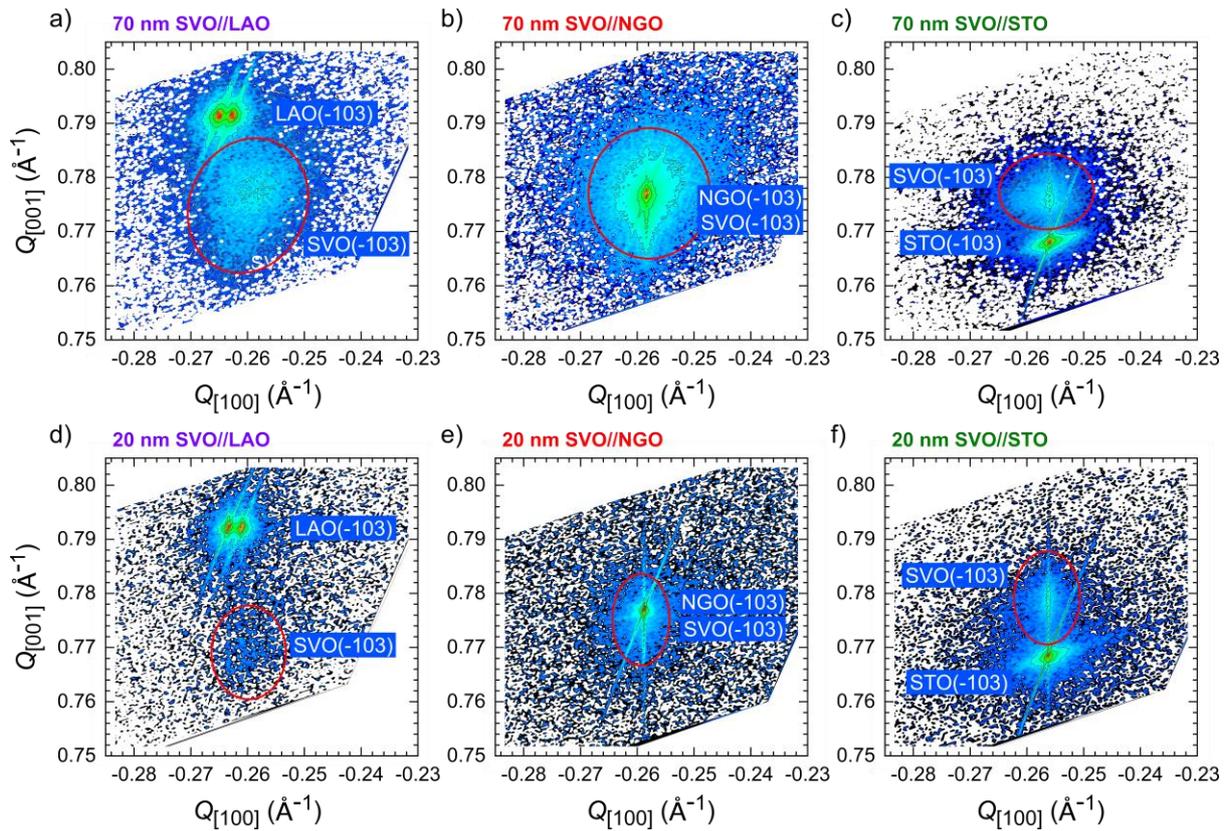

**Figure S2(b):** Corresponding reciprocal space maps measured around the (-103) reflection.

Figure S2(c) below shows a collection of reciprocal space maps measured for the same sample along [100]* and [110]* directions in the reciprocal space. The yellow lines qualitatively indicate the $Q_{[001]}$ position of the substrate and film peaks. Despite the broadening of the film reflections in the out-of-plane direction, because of their reduced thickness (10nm), the centroid positions of the film peaks seem to be independent of the H value of the reflection, both for H03 and HH3 reflections.

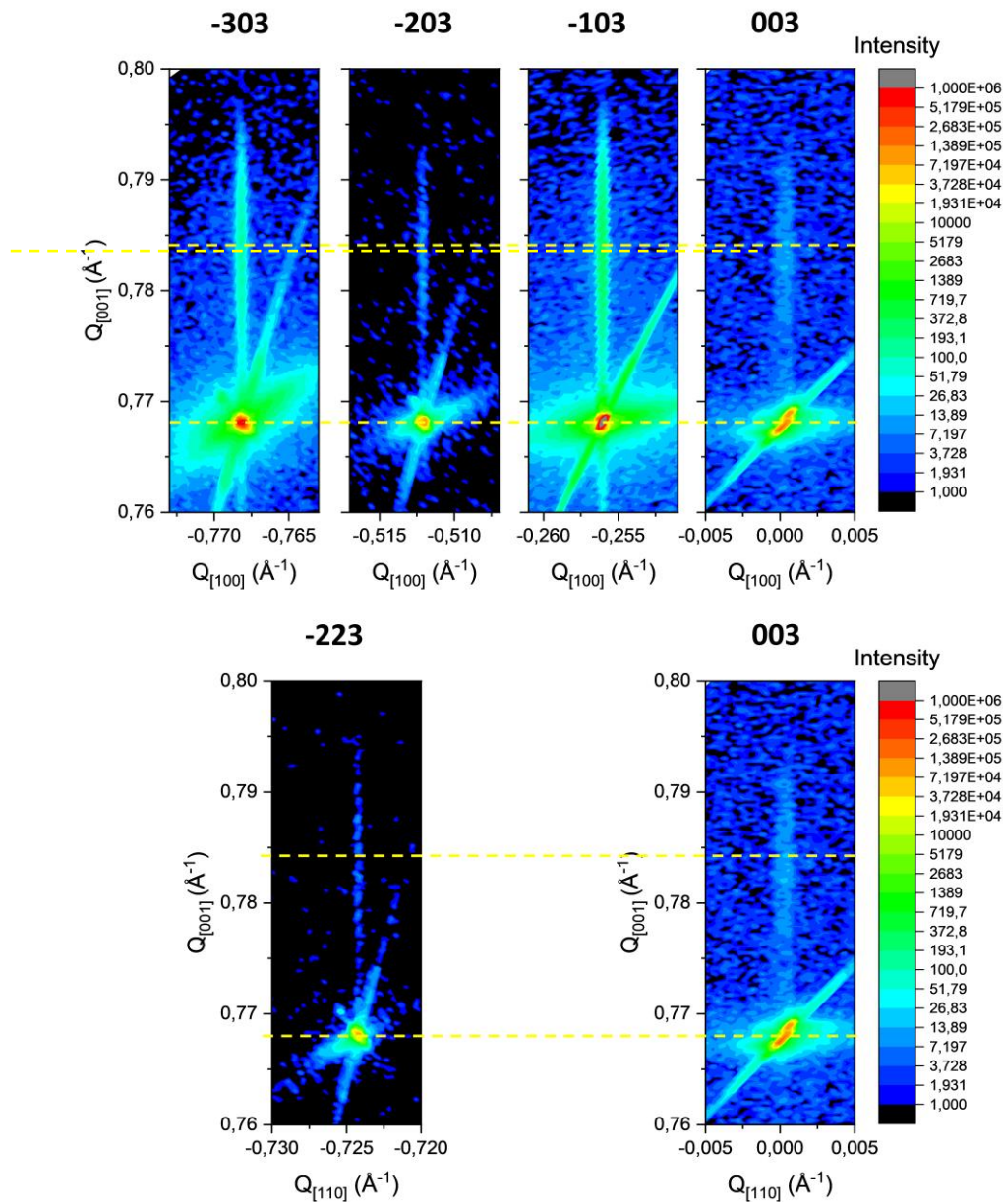

**Figure S2(c):** Reciprocal space maps around selected reflections of the SVO(10 nm)//STO sample. Top panels: H03 reflections (H=0,-1,-2,-3) measured along [100]* direction. Bottom panels: HH3 reflections (H=0, and -2) measured along [110]*.

A more accurate comparison can be obtained by extracting the intensity profiles along the out-of-plane direction for the different reflections as depicted in Figure S2(d). In this representation it is clear that all reflections are centered in the same $Q_{[001]}$ position, and there is no any noticeable splitting or broadening of the peak signal. This is an indication that no significant shear distortion is observed in these films, which points towards an average tetragonal structure of the $SrVO_3$ films.

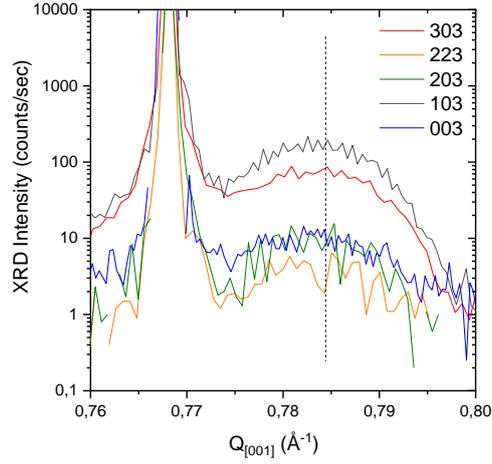

**Figure S2(d):** $Q_{[001]}$ scans of the different H0L and HHL reflections extracted from maps in Figure S2(c).

**Supplementary Information S3**

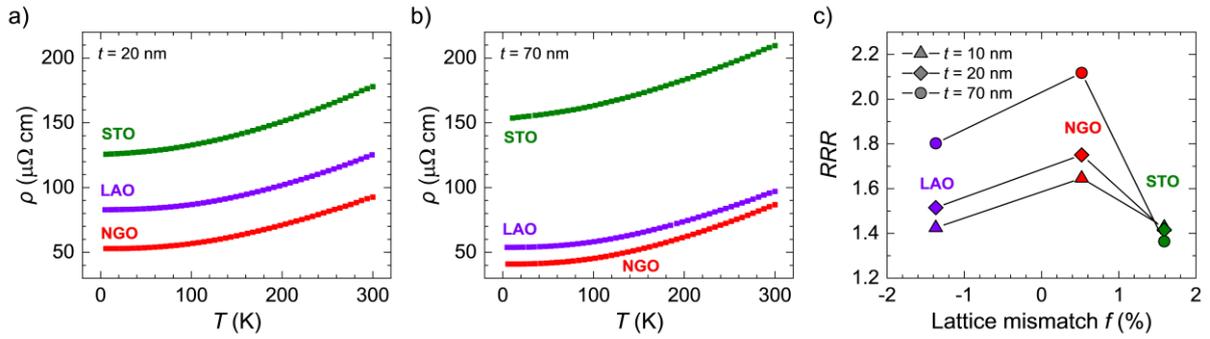

**Figure S3:** (a-b) Temperature-dependent resistivity data of SVO films (20 and 70 nm) deposited at $P(Ar)$ = 0 mbar (*BP*) on various substrates. All films are metallic and the resistivity is smaller on best-matching substrate (NGO). (c) Corresponding residual resistivity ratio ($RRR = \rho(300\ K)/\rho(5\ K)$) for all thicknesses (10, 20, and 70 nm). Films deposited on the best-matching NGO substrate have highest *RRR*, due to a lower presence of planar defects [31].

**Supplementary Information S4**

In Figure S4, we show the *XAS* spectra of both V $L_{2,3}$ and O *K* pre-peak absorption edges for 10 nm SVO//STO sample measured at: a) grazing incidence angle ($\theta = 30°$), and b) a nearly normal incidence angle ($\theta = 85°$) from the sample surface. The spectra were collected for horizontally (*H*) and vertically (*V*) polarized lights and the *XLD* = *H* − *V* is shown in the bottom part of the (a,b) panels. The electric field ***E*** of *H*-polarized light varies from *b*-axis (in-plane) to *c*-axis (out-of-plane) from grazing to normal incidence as shown in the sketches given below the (a,b) panels, respectively. While, the ***E*** of *V*-polarized light is parallel to the *a*-axis, (***E*$_a$** or commonly ***E*$_{ab}$**) for any light incidence angle. It appears from the data that the *XLD* is absent in the normal incidence as expected due to the 3*d*-orbital symmetry in the film plane, but a clear difference in the grazing incidence as the orbital occupancy differs the light absorption intensities. Furthermore, the spectra do not seem to have any discernible shift in the peak positions either at V $L_{2,3}$ or O *K* pre-peak edges - within the experimental resolution (< 10 meV) - suggesting that, within the sensitive penetration depth, the SVO films are electronically homogeneous; this hints that the vanadium valency remains same throughout the film within the x-rays probing limit.

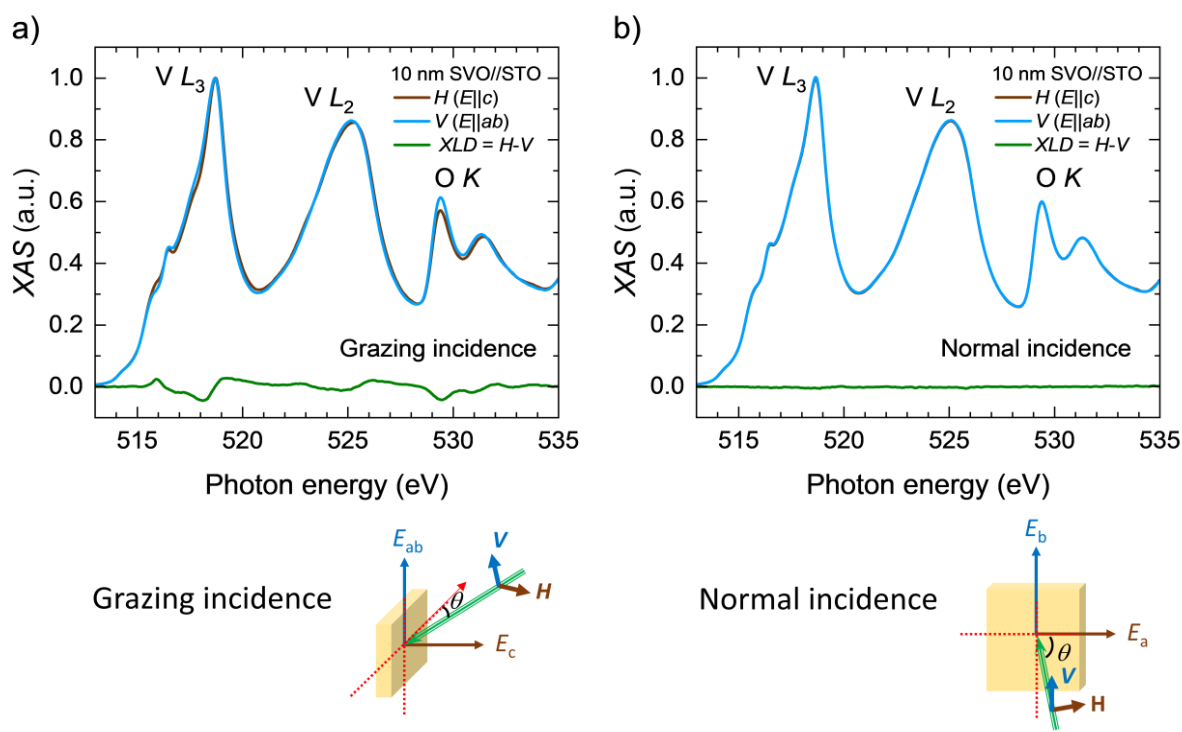

**Figure S4:** *XAS* and *XLD* spectra of V $L_{2,3}$ and O *K* pre-peak for 10 nm SVO//STO recorded at: (a) grazing ($\theta = 30°$), and (b) normal incidence ($\theta = 85°$) angles.

## Supplementary Information S5

In Figure S5(a), we show *H*- and *V*-polarized *XAS* of SVO//STO, as well as *XLD* at the V $L_{2,3}$-edges of 10 nm thick SVO films grown on various substrates (STO, NGO, and LAO) recorded at 2 K temperature. These data were collected at grazing incidence ($\theta$ = 30°) as similar to the 300 K data in the Figure 4(b) of the manuscript. The *XLD-1* and *XLD-2* signals for positive and negative dichroism, respectively, follow the same trend as the 300 K data, i.e. largest amplitude for the SVO films on STO and reduces for the films on NGO and LAO. The magnitudes of *XLD-1* and *XLD-2* are given in the Figure S5(b) for both 2 K and 300 K for a quantitative analysis. The data appear very similar to each other.

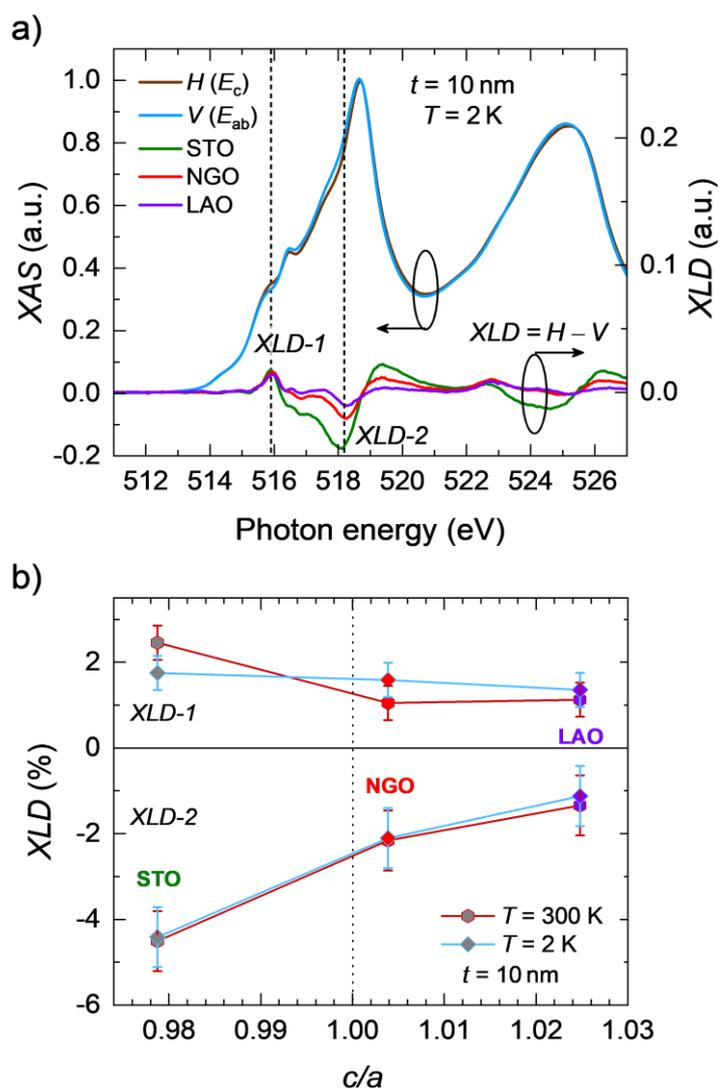

**Figure S5:** (a) *H*- and *V*-polarized *XAS* spectra of SVO//STO at the V $L_{2,3}$-edge, and *XLD* spectra of 10 nm SVO thin films on STO, NGO, and LAO substrates. Spectra were collected at 2 K. (b) Temperature variation of the *XLD* maxima values (*XLD-1* and *XLD-2*) of 10 nm SVO films grown on various substrates.

**Supplementary Information S6**

In Figure S6(a), we show the average *XAS* of SVO films on STO for thicknesses of 10, 20, and 70 nm recorded at grazing incidence, in the energy range of 510-555 eV where the V $L_{2,3}$ absorption edge is present and followed by the O *K*-edge. As already mentioned, the O *K* pre-peak doublet maxima appearing at $M_1$ (≈ 529.5 eV) and $M_2$ (≈ 531.5 eV) (arrows), which are known to be very sensitive to oxygen contents and the hybridization between O 2*p* and metal 3*d*-$t_{2g}$ and between O 2*p* and 3*d*-$e_g$ states, respectively. The relative intensity $I(M_1(t_{2g}))/I(M_2(e_g))$ is sensitive to the electronic occupancy at V 3*d* levels, which seems varying with the increasing thickness due to different surface oxidation in the films. This is very much consistent with the films grown on LAO and NGO for the increasing thicknesses from 10 to 70 nm.

In Figure S6(b-d), we show the *XLD* data for SVO films on STO, NGO, and LAO substrates for different thicknesses (10, 20, and 70 nm) recorded at grazing incidence. The *XLD* displays similar features and the amplitudes at *XLD-1* and *XLD-2* are reduced with the increasing thickness. These *XLD* amplitudes are already summarized in Figure 4(d) of the manuscript.

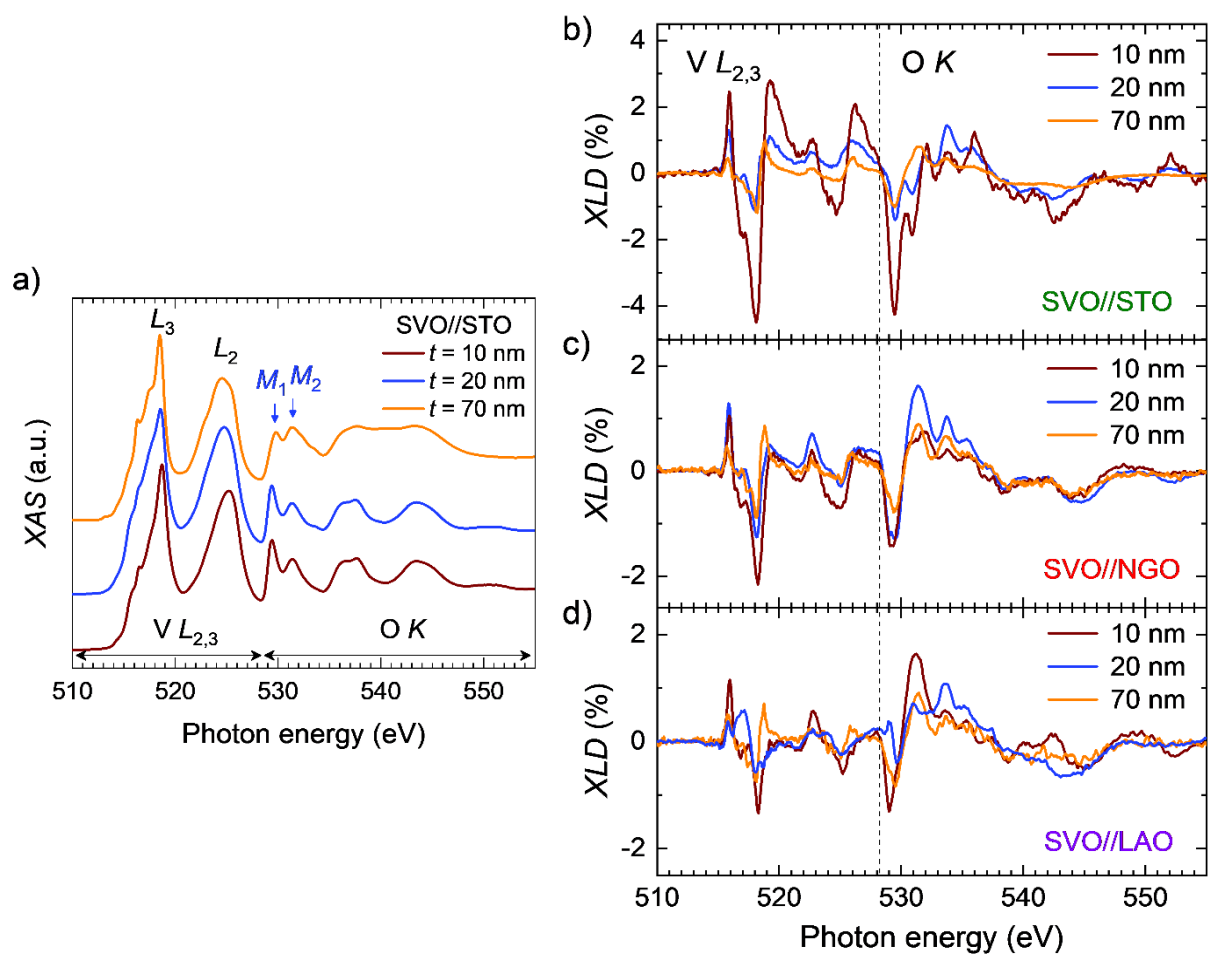

**Figure S6:** (a) The *XAS* of V $L_{2,3}$ and O *K*-edges of SVO//STO for different thicknesses (10, 20, and 70 nm) recorded at the grazing incidence ($\theta$ = 30°). Thickness variation of *XLD* spectra of V $L_{2,3}$ and O *K*-edges for the SVO films on: (b) STO, (c) NGO, and (d) LAO substrates recorded at the grazing incidence.

**Supplementary Information S7**

In Figure S7, we show the *XLD* data of V $L_{2,3}$ and O *K*-edges recorded at different angles of incidence from the sample surface (8°, 30°, 60° and 85°) for the 10 nm SVO films on: (a) STO, (b) NGO, and (c) LAO substrates. These data confirm the systematic variation of the *XLD* signal with the substrate as well as the angle. In Figure S7(d) we summarize the amplitudes of the *XLD* maxima (*XLD-1* and *XLD-2*) for various substrates. The *XLD* maxima are plotted as a function of $\cos^2(\theta)$ to provide evidence of the concordance of the expected angular dependence with experimental data.

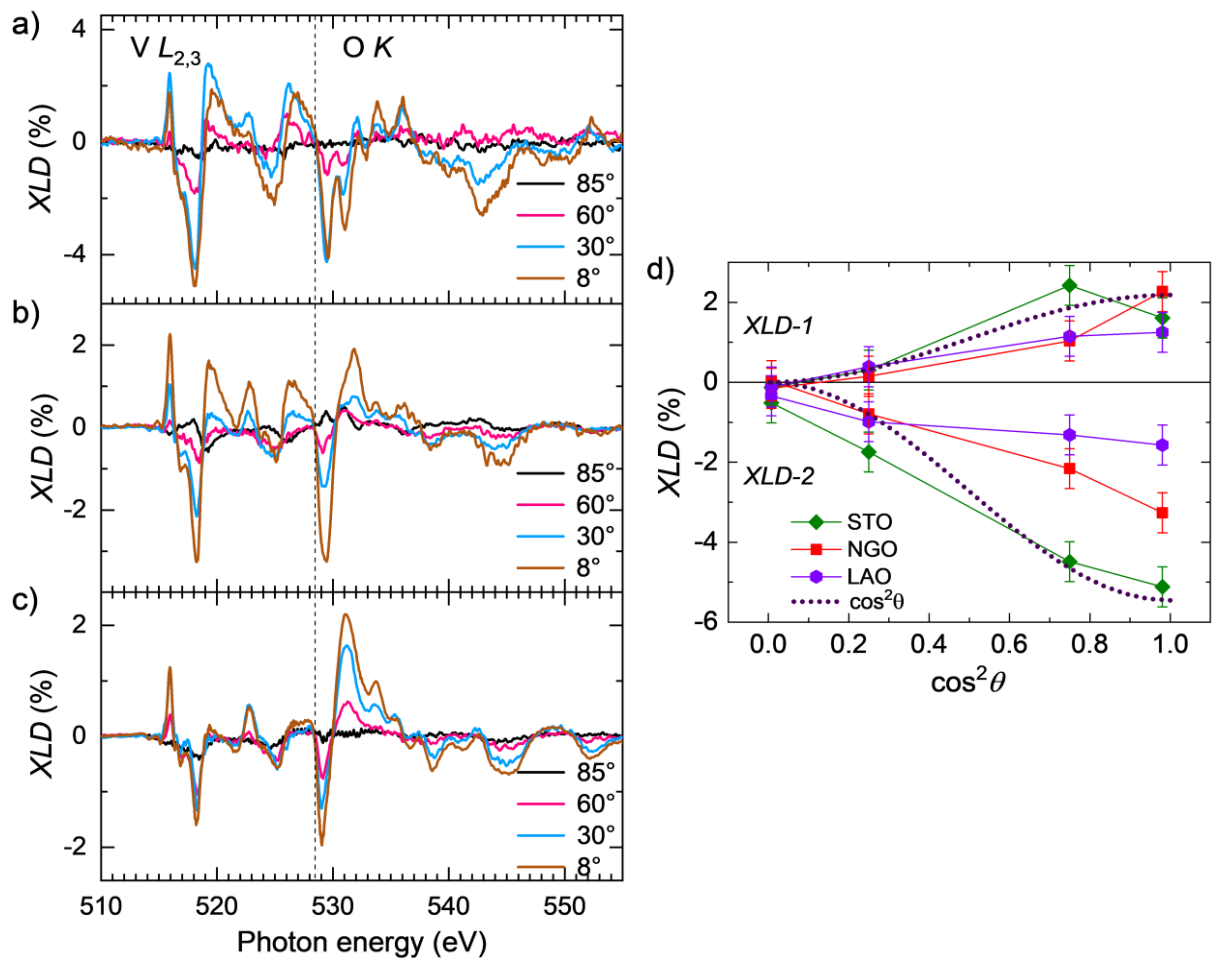

**Figure S7:** (a-c) Angular dependence of the *XLD* signals V $L_3$-edge for 10 nm SVO films on different substrates (STO, NGO, and LAO). (d) Summary of the *XLD-1* and *XLD-2* maxima amplitudes for the different angles.

**Supplementary Information S8**

It can be appreciated in Figure S8 that: (i) the occupancy of $t_{2g}$ is larger than $e_g$ as expected from the crystal field splitting of octahedral $VO_6$ coordination; (ii) the in-plane $xy$ states are more occupied than $xz, yz$ (which are degenerate) when $c/a < 1$, illustrating that they are pushed down under tensile strain; the opposite trend holds for out-of-plane ($xz, yz$) when $c/a > 1$; (iii) surprisingly the in-plane $x^2-y^2$ states display the opposite behavior, that is: they appear to be less occupied under tensile strain ($c/a < 1$), and consistently, the $z^2$ states become less occupied under compressive strain ($c/a > 1$). To explain this behavior it is helpful to calculate the center of mass $E_{CM}$, i.e. the average energy at which the $d$ orbitals are located. Without any bonding-antibonding hybridization the *DOS* would be located around $E_{CM}$. When bonding-antibonding hybridization happens, the sign of $E_F - E_{CM}$ determines whether the bonding or the antibonding states have the dominant contribution below $E_F$. For the $t_{2g}$ states $E_{CM} < E_F$, therefore the antibonding states, being partially occupied, control the behavior of the *IDOS*, as already explained in (i). For the $e_g$ states, the antibonding states are unoccupied and lie at high energies and, as expected, $E_{CM} > E_F$, which results in the bonding states to be the only contribution to the *IDOS* below $E_F$. With respect to $E_F$ the bonding states act exactly opposite to the antibonding states, and therefore, the IDOS for the $e_g$ orbitals follows the opposite trend to the *IDOS* for $t_{2g}$ states, as depicted in Figure S8.

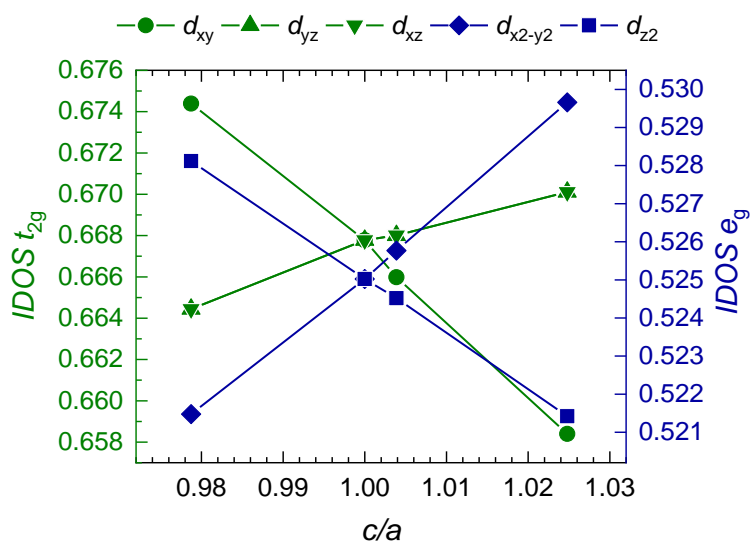

**Figure S8:** *IDOS* of $t_{2g}$ (left) and $e_g$ (right) *vs* tetragonal distortion ($c/a$), for in-plane and out-of-plane orbitals, as indicated.

**Supplementary Information S9**

Charge unbalance at $d_{xy}$, $d_{xz}$, $d_{yz}$ orbitals and $p_x$, $p_y$, $p_z$ as inferred from the analysis of the XLD data.

Under strain, ignoring octahedral rotation and considering only tetragonal deformations of the SVO lattice, for a fixed amount of charge at $t_{2g}$ levels ($d^1$), the sign of XLD of $V^{4+}$ can be expected to change its sign when going from tensile to compressive strain.

However, this statement neglects the contribution of surface. We had earlier discovered [14,18] that surface symmetry breaking unavoidably favors occupancy of out-of-plane orbitals and demonstrated that this contribution can preclude the observation of a change of sign of XLD with strain. One can write: XLD = XLD(s)+XLD(b), where XLD(s) refers to the surface contribution (always of a given sign) and XLD(b) refers to the contribution of bulk, that can change with the sign of strain. Observation of a change of sign in the experimental XLD is determined by the relative weight of both contributions. In the present case, detailed inspection of data allowed us to exclude surface contribution as ruling the observed XLD($c/a$) variation.

We were then faced by the possible role of the 2p orbitals and their covalent mixing with $t_{2g}$. Figure 8(a) suggests that the $t_{2g}$ orbitals – taken alone – would induce somehow a change of sign of XLD from $c/a > 1$ to $c/a < 1$. However, Figure 8(c) shows that $d/p$ (in-plane) density of states ratios and the corresponding $d/p$ (out-of-plane) both change with $c/a$, implying the charge is transferred within the $p$-$d$ hybrid orbitals. Charge at V ions is thus not preserved and the simplest picture of changing of sign of XLD, obtained by assuming a fixed number of carriers at V orbitals, no longer holds. Charge redistribution involving ($t_{2g}$, $e_g$)-2$p$ orbitals occurs.

This is the message that we aim to convey in this manuscript. The sketch below (Figure S9(a)) illustrates that according to the XLD analysis, as described in main text.

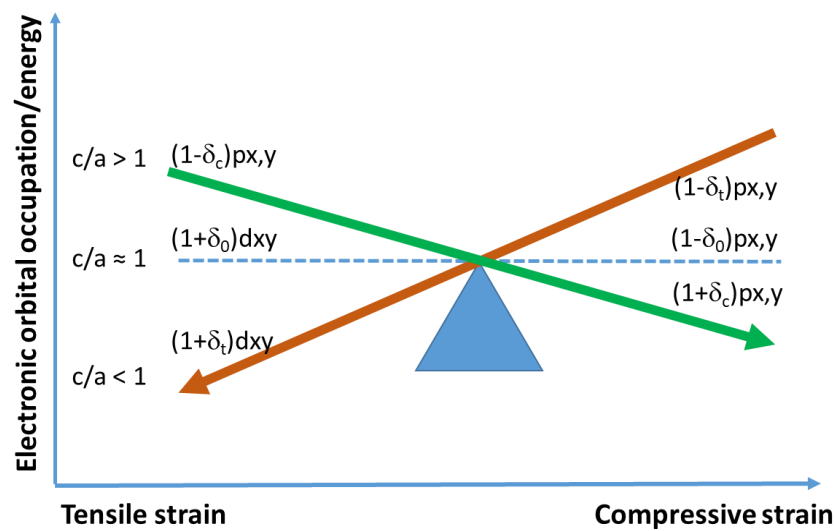

**Figure S9(a):** Change of the relative occupation (or energy position) of $d_{xy}$ orbitals vs $p_{x,y}$ as a function of the tetragonality ratio $c/a$. The relative weight of the hybrid orbitals are indicated by

$(1+\delta_j)d_{xy}+(1-\delta_j)p_{x,y}$ where $\delta_j$ (j = 0, t, c) represents the electron transfer for unstrained (j = 0) film, compressive strain (j = c) and tensile strain (j = t).

The trends sketched above can precisely observed in the DFT data shown in the manuscript. To allow a simpler view we have plotted here the integrated density of states (IDOS) for the $d_{xy}$ and $p_x$, $p_y$ orbitals for the STO and LAO cases, representing the limiting cases of tensile and compressive strain explored here. At zero temperature, the energy integration from an isolated set of (**k**-integrated) bands up to the Fermi level $E_F$ provides information of the occupied electron states. The resulting Figure S9(b) below verifies the sketch in Figure S9(a) derived from the experimental data where the change in hybridization is reflected in a change of relative electron occupation between $d$ and $p$ orbitals.

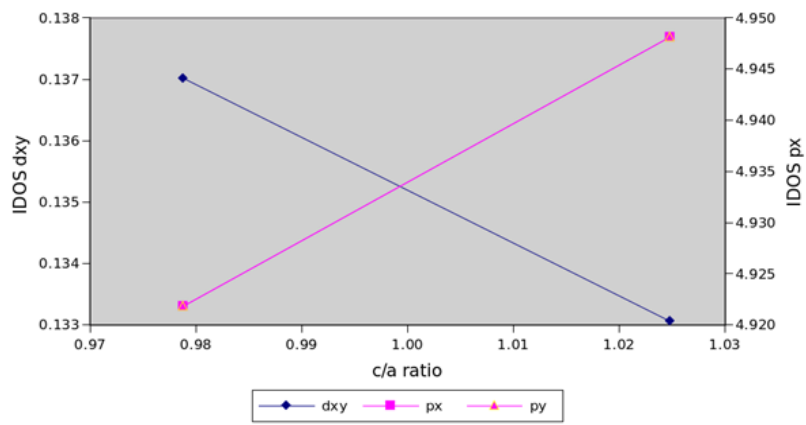

**Figure S9(b):** Integrated density of states vs. $c/a$ ratio. The left axis describes the change of IDOS for $d_{xy}$ (blue line) and the right axis the change of $p_x$ and $p_y$ (pink lines). This figure corresponds to the sketch in Figure S9(a) showing the redistribution of the relative weight for $d_{xy}$ and $p_x$, $p_y$.

Furthermore, by visualizing the orbitally resolved DOS directly, one can clearly observe a difference in the position of bonding and antibonding states, corresponding directly to a change of hybridization, please see Figure S9(c).

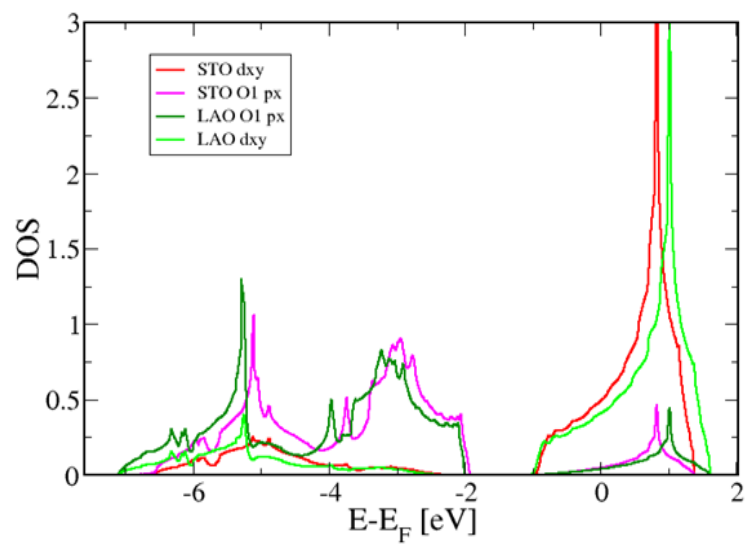

**Figure S9(c):** Density of states vs. energy with respect to the fermi level. For clarification only the $d_{xy}$ and O1 $p_x$ orbital are depicted. Visible are bonding and antibonding states for both orbitals for STO and LAO. In dependence of the $c/a$ ratio the distance of bonding and antibonding changes, which goes hand in hand with a change of hybridization.